\documentclass[preprint,12pt]{aastex}
\usepackage{graphicx}
\usepackage{amsmath,amssymb}
\usepackage{natbib}
\usepackage{array}
\usepackage{mypack}
\usepackage{longtable}
\usepackage[caption=false]{subfig}
\usepackage{lscape}

\begin{document}

\title{Pulsar J0453+1559: A Double Neutron Star System with a Large Mass Asymmetry}

\author{J.G. Martinez$^{1,3}$, K. Stovall$^{1,2}$, P.C.C. Freire$^{3}$,
  J.S. Deneva$^{4}$, F.A. Jenet$^{1}$, M.A. McLaughlin$^{5}$,
  M.Bagchi$^{6,5}$, S.D. Bates$^{5}$, A. Ridolfi$^{3}$}

\footnotetext[1]{Center for Advanced Radio Astronomy, University of
  Texas at Brownsville, One West University Boulevard, Brownsville, TX
  78520}
\footnotetext[2]{Department of Physics and Astronomy, University of
  New Mexico, Albuquerque, New Mexico 87131}
\footnotetext[3]{Max-Planck-Institut f\"{u}r Radioastronomie, Auf dem
  H\"{u}gel 69, D-53121 Bonn, Germany}
\footnotetext[4]{National Research Council, resident at the Naval 
  Research Laboratory, Washington, DC 20375}
\footnotetext[5]{Department of Physics and Astronomy, West Virginia
  University, 111 White Hall, Morgantown, WV 26506}
\footnotetext[6]{The Institute of Mathematical Sciences, 4th Cross
  Road, CIT Campus Taramani, Chennai 600 113, India}

\begin{abstract}

To understand the nature of supernovae and neutron star (NS) formation, 
as well as binary stellar evolution and their interactions, it is
important to probe the distribution of NS masses. Until now, all 
double NS (DNS) systems have been measured to have a mass ratio 
close to unity ($q\, \geq\, 0.91$). Here we report the measurement of the individual 
masses of the 4.07-day binary pulsar J0453+1559
from measurements of the rate of advance of periastron and Shapiro delay:
The mass of the pulsar is $M_{p} = 1.559  \, \pm\, 0.005 \, \Msun$  and that of
its companion is $M_{c} \, = \, 1.174 \, \pm \, 0.004 \, \Msun$;
$q = 0.75$.
If this companion is also a neutron star (NS), as indicated by
the orbital eccentricity of the system ($e = 0.11$), then its mass is the
smallest precisely measured for any such object.
The pulsar has a spin period of 45.7 ms and a spin period derivative 
$ \dot{P} = (1.8616\, \pm \, 0.0007) \,\times\, 10^{-19} \rm \, s~s^{-1}$; from these we derive a
characteristic age of $\sim 4.1\, \times \, 10^9$ years and a magnetic
field of $\sim \, 2.9 \, \times \, 10^{9}$ G, i.e,
this pulsar was mildly recycled by accretion of matter from the
progenitor of the companion star. This suggests that it
it was formed with (very approximately) its current mass. Thus NSs
form with a wide range of masses, which is important for
understanding their formation in supernovae. It is also important for 
the search for gravitational waves released during a NS-NS merger: 
it is now evident that we should not assume all DNS systems are symmetric.
\end{abstract}

\maketitle

\section{Introduction}
Double neutron star (DNS) systems are rare and valuable physical
laboratories that can be used to precisely test gravity theories. The
first such system, PSR B1913+16, provided evidence for orbital decay
due to the emission of gravitational waves as predicted by general
relativity (GR, \citealt{ht75} and \citealt{htNobel}). Since the
discovery of PSR B1913+16, nine additional DNS systems have been
discovered in the Galaxy (see Table~\ref{table:DNSlist}), including one such system
in which both neutron stars (NSs) have been detected as radio pulsars,
PSRs J0737$-$3039A and B \citep{doublePSR}. This system
provides one of the best available tests of GR and alternative
theories of gravity in the strong-field regime
\citep{GRTestsdoublePSR}. 

DNS systems begin as two high-mass stars. The higher-mass star will
undergo a supernova explosion resulting in a neutron star and a
high-mass companion. Prior to the supernova of the companion, there is
typically a period of mass transfer from the companion onto the
neutron star and the system can be detected as a high-mass X-ray
binary. Eventually, the companion will undergo a supernova explosion,
leaving behind two neutron stars: the older might be detected as a
mildly recycled pulsar which was spun up by accretion from the
progenitor of the younger star, the younger might be detected as a
normal pulsar. In the rare case that the system survives both
supernovae, the result is a DNS (see \citep{DuncReview} and references
therein). 

In this Letter, we report the timing solution for
PSR J0453+1559, a pulsar discovered in the Arecibo 327 MHz Drift Pulsar
Survey \citealt{DenevaAODrift}. At the time of writing has discovered
this survey has discovered a total of 62 pulsars and rotating radio transients.
As reported by \citealt{DenevaAODrift}, PSR J0453+1559 has a
spin period of 45.7 ms, a dispersion measure (DM) of 30.3
pc~cm$^{-3}$, an orbital period of 4,07 days and a massive companion.

In section 2, we describe the observations, data reduction and the
derivation of the timing solution and in section 3, we present the
timing parameters of this new system. In section 4 we conclude with a
discussion of the significance of the timing parameters.

\begin{landscape}
\centering
\begin{table}[h!]
\begin{footnotesize}
\centering
\caption{Double neutron star systems known in the Galaxy}
\begin{tabular}{l r@{.}l r@{.}l r@{.}l l l r@{.}l r@{.}l c}
\hline\hline
Pulsar & \multicolumn{2}{l}{Period}  &  \multicolumn{2}{l}{$P_\mathrm{b}$} &  \multicolumn{2}{l}{$x_\mathrm{}$} & $e_\mathrm{}$ & $M$ &  \multicolumn{2}{c}{$M_\mathrm{p}$} &  \multicolumn{2}{c}{$M_\mathrm{c}$} & References \\
 & \multicolumn{2}{l}{(ms)} &  \multicolumn{2}{l}{(days)} &  \multicolumn{2}{l}{(lt-sec)} & & ($\Msun$) &  \multicolumn{2}{c}{($\Msun$)} &  \multicolumn{2}{c}{($\Msun$)}  \\
%heading
\hline
J0737$-$3039A & 22&699   &     0&102    &   1&415  &  0.0877775(9)  & 2.58708(16) &   1&3381(7) &   1&2489(7) & 1 \\
J0737$-$3039B & 2773&461 &    \multicolumn{2}{l}{} & 1&516   \\
J1518+4904    & 40&935   &    8&634     &  20&044  & 0.24948451(3)  & 2.7183(7)   &     \multicolumn{2}{l}{-} &  \multicolumn{2}{l}{-}   &   2 \\
B1534+12      & 37&904   &     0&421    &   3&729  & 0.27367740(4)  & 2.678463(4) &   1&3330(2) &   1&3454(2) & 3 \\
J1753$-$2240  & 95&138   &    13&638    &  18&115  & 0.303582(10)   & \multicolumn{2}{l}{-}  & \multicolumn{2}{l}{-}  & \multicolumn{1}{l}{-} & 4 \\
J1756$-$2251  & 28&462   &    0&320     &   2&756  & 0.1805694(2)   & 2.56999(6)  &   1&341(7)  &   1&230(7)  & 5 \\
J1811$-$1736  & 104&1    &   18&779     &  34&783  & 0.82802(2)     & 2.57(10)    &     \multicolumn{2}{l}{-} &  \multicolumn{2}{l}{-}    & 6 \\
J1829+2456    & 41&009   &    1&760     &   7&236  & 0.13914(4)     & 2.59(2)      &    \multicolumn{2}{l}{-} &  \multicolumn{2}{l}{-}    & 7 \\
J1906+0746*    &144&073   &    0&166     &   1&420  & 0.0852996(6)   & 2.6134(3)   &   1&291(11) &   1&322(11) & 8 \\
B1913+16      & 59&031   &    0&323     &   2&342  & 0.6171334(5)   & 2.8284(1)   &   1&4398(2) &   1&3886(2) & 9 \\
J1930$-$1852  & 185&520  &   45&060     &  86&890  & 0.39886340(17) & 2.59(4)     &     \multicolumn{2}{l}{-} &  \multicolumn{2}{l}{-}    & 10 \\
{\bf J0453+1559}& 45&782 &    4&072     &  14&467  & 0.11251832(4)  & 2.734(3)    &   1&559(5)  &   1&174(4)  & This Letter\\
\hline
 \multicolumn{14}{l}{Globular cluster systems} \\
\hline
J1807$-$2500B* & 4&186    &    9&957     &  28&920  & 0.747033198(40) & 2.57190(73) &  1&3655(21) & 1&2064(20)   & 12 \\
B2127+11C     & 30&529   &    0&335     &   2&518  & 0.681395(2)    & 2.71279(13) &   1&358(10) &  1&354(10)   & 13 \\
\hline
\tablecomments{1:~\cite{doublePSR} \& ~\cite{GRTestsdoublePSR}, 2:~\cite{JanssenJ1518+4904}, 3:~\cite{B1534} \& ~\cite{B1534_updated}, 4:~\cite{J1753}, 5:~\cite{J1756} \& ~\cite{J1756_updated}, 6:~\cite{CorongiuJ1811-1736}, 7:~\cite{J1829} \& ~\cite{J1829_2}, 8:~\cite{J1906} \& ~\cite{J1906_updated}, 9:~\cite{ht75} \& ~\cite{Weisberg}, 10:~\cite{J1930}, 12:~\cite{J1807}, 13:~\cite{B2127} \& ~\cite{B2127_updated} 
\\
* Note: there is some uncertainty on whether these systems are DNSs. }
\end{tabular}
\label{table:DNSlist}
\end{footnotesize}
\end{table}
\end{landscape}

\newpage

\section{Observations and data reduction}
PSR~J0453+1559 was observed with the L-wide receiver of the 305-m
Arecibo radio telescope 45 times over 2.5 years using the Puerto
Rico Ultimate Pulsar Processing Instrument (PUPPI, a clone of the
Green Bank Ultimate Pulsar Processing Instrument, GUPPI)\footnote{
  http://safe.nrao.edu/wiki/bin/view/CICADA/GUPPISupportGuide} as a
back-end, which allows simultaneous processing of the
 $\Delta f = 600$ MHz bandwidth provided by the receiver
(from 1130 to 1730 MHz) with a system temperature
$T_{\rm sys} = 30$ $\rm K$ and a gain $G = 10$ $\rm K/Jy$. 
The first six months' observations were taken in search mode
with 2048 channels,
a time resolution $T_{\rm res}$ of 40.96 $\mu $s, 
two polarizations ($n_{p} = 2$) and basically no sensitivity degradation
due to  digitization ($\beta \simeq 1$), since PUPPI digitizes the antenna
voltages with 8 bits. After derivation of a phase-connected
timing solution, these were dedispersed and folded using
{\tt DSPSR}\footnote{http://dspsr.sourceforge.net/},(\citealt{dspsr}),
producing 1024-bin profiles.

\begin{figure}[h!]
\begin{center}
\includegraphics[width=4.5in, angle=270]{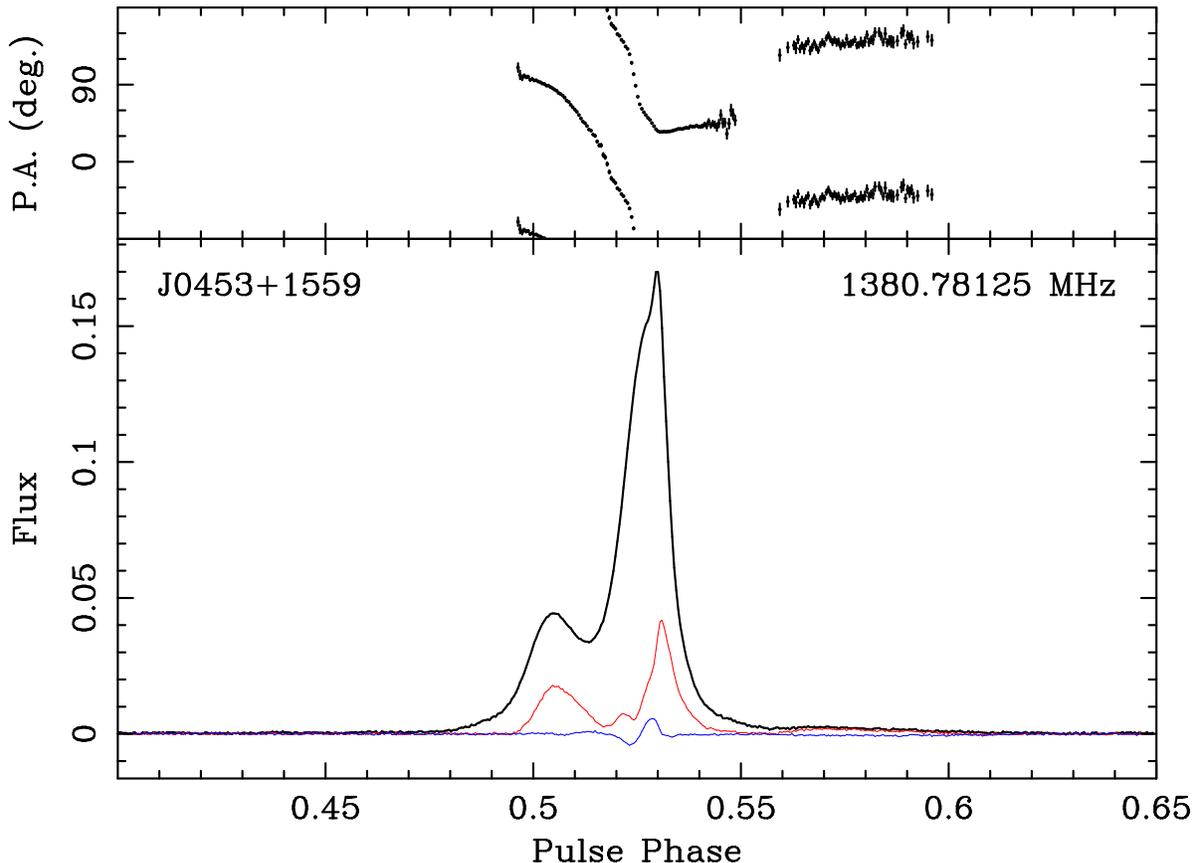}
\end{center}
\caption [] {Pulse profile for PSR J0453+1559 at L-band (1170-1730 MHz),
obtained by averaging the best detections of the pulsar.
The black line indicates total intensity, the red line is the amplitude
of linear polarization and the blue line is the amplitude of the circular
polarization. In the top panel, we depict the position angle of the
linear polarization, where a clear polarization swing and a sudden jump
between orthogonal modes is clearly visible.
}
\label{figure:profile}
\end{figure}

This initial timing ephemeris made it possible to conduct
all subsequent observations in coherent fold mode (with 512 channels, 2048 
phase bins and 4 Stokes parameters), which coherently
dedisperses and folds the data online, optimally removing the
dispersive effects of the interstellar medium. These observations have
improved signal-to-noise ratio because they benefit for the
better pointing position derived from the timing solution. The
pulse profile is displayed in Figure~\ref{figure:profile}. The main
pulse has a sharp feature that contributes much to the good timing
precision of this pulsar discussed in section 3.

The dedispersed pulse profiles obtained when averaging the 11-minute blocks 
of timing data produced by PUPPI are then calibrated using the noise diode
observations that are taken with (almost) every single observation. 
The resulted calibrated pulse profiles are then cross-correlated
with the low-noise template displayed in Fig.~\ref{figure:profile} 
using the procedure described in \cite{TaylorRG} and implemented in
the {\tt PSRCHIVE} software (\citealt{PSRCHIVE}, \citealt{PSRCHIVE2012}).
This resulted in 868 usable topocentric pulse times of arrival (TOAs).

We then used {\tt TEMPO}\footnote{http://tempo.sourceforge.net/}
to correct the TOAs using the Arecibo telescope's clock corrections
and to convert them to the Solar System barycentre. To do this, the
motion of the radio telescope relative to the Earth was calculated
using the data from the International Earth Rotation Service, and to
the barycenter using the DE421 solar system
ephemeris \footnote{ftp://ssd.jpl.nasa.gov/pub/eph/planets/ioms/de421.iom.v1.pdf}.
Finally, the difference between the measured TOAs and those predicted
by a model of the spin and the orbit of the pulsar is minimized by
{\tt TEMPO}, by varying the parameters in the model. The parameters
that best fit the data are presented in the first column of Table 
\ref{table:0453}. To model the orbit, we used the DDGR model described by
\cite{Damour85} and \cite{Damour86}, which assumes the validity of GR
in the description of the orbital motion of the system and uses as
parameters the total mass of the system $M$ and the companion mass
$M_c$.

\begin{figure}[h!]
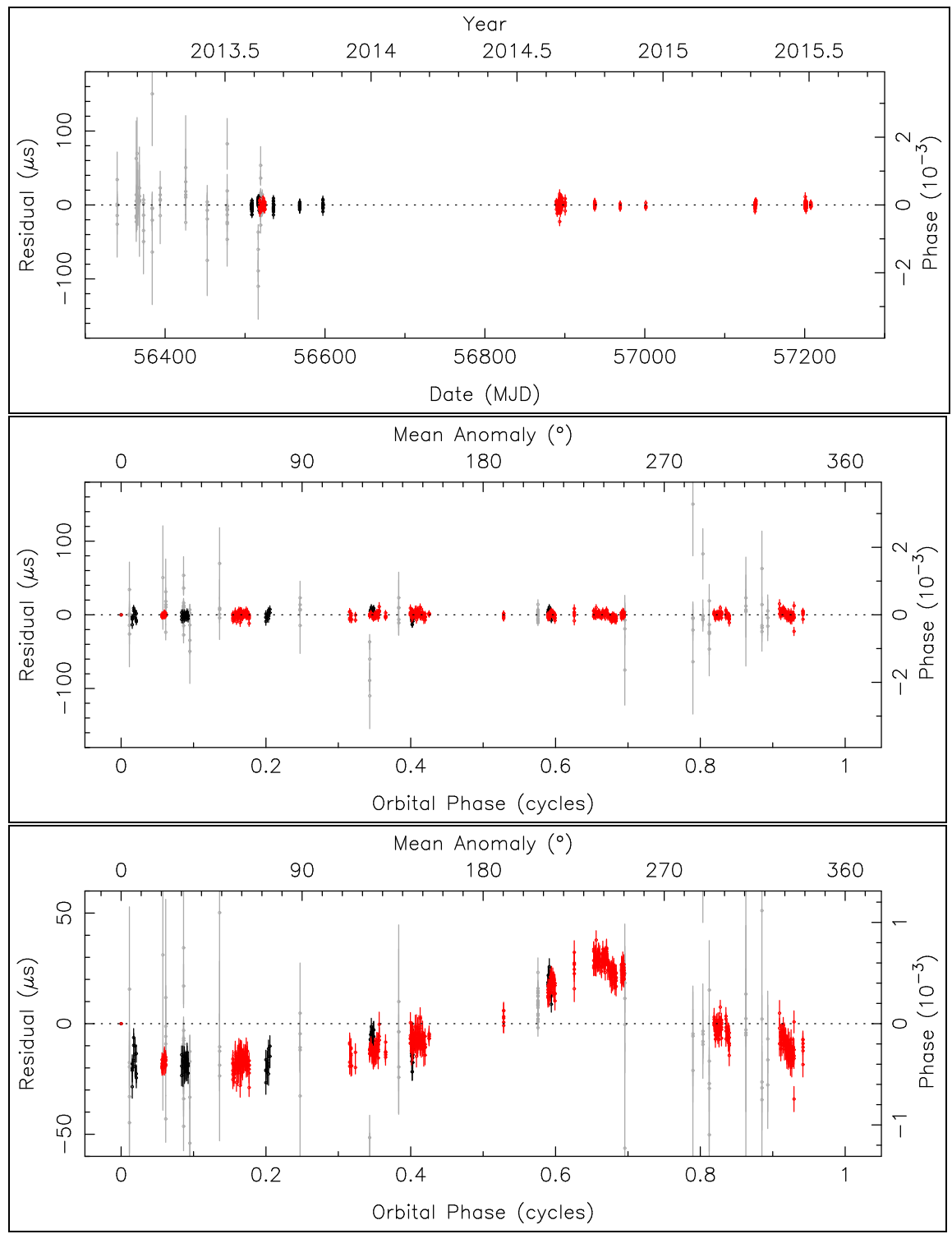

\label{figure:TOAs}
\fbox{\includegraphics[width=5in, angle=270,scale=0.5]{Residuals_vs_Epoch.ps}}
\hspace{5px}
\fbox{\includegraphics[width=5in, angle=270,scale=0.5]{Residuals_vs_Phase.ps}}
\hspace{5px}
\fbox{\includegraphics[width=5in, angle=270,scale=0.5]{Residuals_vs_phase_no_Shapiro.ps}}
\hspace{5px}
\caption{\small The top plot is the timing solution for PSR J0453+1559, timing residuals (measured pulse arrival times - model pulse arrival times) as a function of MJD. The middle and bottom plot show the timing residuals versus orbital phase of the J0453+1559 system. The bottom plot shows the magnitude of the Shapiro Delay as a function of orbital phase, derived with the same Keplerian orbital parameters from the DDGR ephemeris in Table~\ref{table:0453}. The color of the timing residuals are categorized by: gray is search mode data, black is coherent fold mode that is not calibrated data, and red is coherent fold mode calibrated data. }
\end{figure}

The residuals (TOAs minus model predictions) associated with this DDGR
model are displayed in Figure~\ref{figure:TOAs}. There are some short term
trends in the residuals that point towards unmodeled systematics. For
this reason we added 2.5 $\mu$s (the approximate amplitude of
these systematics) in quadrature to the TOA uncertainties,
in this way the reduced $\chi^2$ is close to 1.0 both for TOAs with
large and small uncertainties. The amplitude of the systematics is smaller
for the data with polarization calibration, which suggests imperfect polarization
calibration might be a cause of the systematics in the uncalibrated data.
The residual root mean square is 4 $\mu$s,
which represents a fraction of $8.7 \times 10^{-5}$ of the spin period. 

In order to double check the results, we used {\tt TEMPO2} (\citealt{tempo2} \& 
  \citealt{tempo22}) and the DDH
model described by \cite{FreireWex}, which like the DD (but unlike the
DDGR model) allows a theory-independent fit for the detectable
post-Keplerian (PK) parameters: in this case the rate of advance of
periastron ($\dot{\omega}$) and two parameters that provide an
optimized description of the Shapiro delay, the orthometric amplitude
($h_3$) and the orthometric ratio ($\varsigma$). As
we will see, the measurement of the PK parameters is important because
it allows an understanding of the precision of the mass measurements
derived from the DDGR model and also a double check on its accuracy.

\begin{table}[h!]
  \begin{center}{\scriptsize
  \caption{}
  \begin{tabular}{l c c}
  \hline
  \multicolumn{3}{c}{Timing Parameters for PSR J0453+1559}\\
  \hline\hline
  Fitting program \dotfill & TEMPO & TEMPO2 \\
  Time Units \dotfill & TDB & TCB \\
  Solar system ephemeris \dotfill & DE421 & DE421 \\
  Reference Epoch (MJD) \dotfill & 56400 & 56400 \\
  Span of Timing Data (MJD) \dotfill & 56339-57207 & 56339-57207 \\
  Number of TOAs \dotfill &  868 & 868 \\
  RMS Residual ($\us$) \dotfill & 3.96 & 3.88 \\
  Solar $n_0$ (cm$^{-3}$) \dotfill & 0.0 & 0.0 \\
  Right Ascension, $\alpha$ (J2000) \dotfill & 04:53:45.41372(4) & 04:53:45.41368(5) \\
  Declination, $\delta$ (J2000) \dotfill & +15:59:21.3055(50) & +15:59:21.3063(59) \\
  Proper motion in RA, $\mu_{\alpha}$ (mas yr$^{-1}$) \dotfill & $-$5.4(4) & $-$5.5(5) \\
  Proper motion in DEC, $\mu_{\delta}$ (mas yr$^{-1}$) \dotfill & $-$5.3(3.6) & $-$6.0(4.2)  \\
  Pulsar Period, $P$ ($\s$) \dotfill & 0.045781816163093(3) &0.0457818168729515(33)   \\
  Period Derivative, $\dot{P}$ ($10^{-18} \rm s\, \ps$) \dotfill & 0.18616(7) &1.8612(8) \\
  Dispersion Measure, DM (\dmu) \dotfill & 30.30527(26) & 30.3053(3) \\
  \hline
  \multicolumn{3}{c}{Binary Parameters}\\
  \hline\hline
  Orbital model \dotfill & DDGR & DDH \\
  Orbital Period, $P_b$ (days) \dotfill & 4.072468588(4) & 4.072468649(4) \\
  Projected Semi-major Axis of the pulsar orbit, $x$ (lt-s) \dotfill & 14.466798(5) & 14.4667896(42) \\
  Epoch of Periastron, $T_0$ (MJD) \dotfill & 56344.0029907(6) & 56344.0031965(9) \\
  Orbital Eccentricity, $e$ \dotfill & 0.11251832(4) & 0.11251844(8) \\
  Longitude of Periastron, $\omega$ ($^\circ$) \dotfill & 223.06953(6) & 223.06965(8) \\
  \hline
  \multicolumn{3}{c}{Relativistic parameters and masses}\\
  \hline\hline
  Rate of advance of periastron, $\dot{\omega}$ ($^\circ \rm yr^{-1}$) \dotfill & 0.0379412 (d) & 0.03793(3) \\
  Orthometric amplitude, $h_3$ ($\mu$s) \dotfill & - & 3.07(25) \\
  Orthometric ratio, $\varsigma$ \dotfill & - & 0.709(40) \\
  Total Mass, $M$ (\Msun) \dotfill & 2.734(4) & 2.733(4) (d) \\
  Companion Mass, $M_{c}$ (\Msun) \dotfill & 1.174(4) & 1.172(4) (m) \\
  \hline
  \multicolumn{3}{c}{Derived parameters}\\
  \hline\hline 
  Mass function, $f$ (\Msun) \dotfill & 0.19601284(22) &0.19601248(4)   \\
  Pulsar mass, $M_{p}$ (\Msun) \dotfill & 1.559(5) & 1.560(5) (m) \\
  Orbital inclination, $i$ ($^\circ$)  \dotfill & 75.2699 & $75.7^{+0.7}_{-0.8}$ (m)\\
  Galactic Longitude, $l$ \dotfill &  \multicolumn{2}{c}{184.1245}\\
  Galactic Latitude, $b$ \dotfill & \multicolumn{2}{c}{$-$17.1369}\\
  DM Derived Distance, $d$ (kpc) \dotfill & \multicolumn{2}{c}{1.1}\\
  Galactic height, $z$ (kpc) \dotfill & \multicolumn{2}{c}{$-$0.29}\\
  Transversal velocity, $v_{\perp}$ (km s$^{-1}$) \dotfill &
  \multicolumn{2}{c}{ $40$}\\
  Kinematic correction to $\dot{P}$, ($10^{-18} \rm s\, \ps$) \dotfill  & \multicolumn{2}{c}{ $0.0089$ }\\
  Intrinsic $\dot{P}$, ($10^{-18} \rm s\, \ps$) \dotfill & \multicolumn{2}{c}{ $\simeq$0.177(2) }\\
  Surface Magnetic Field Strength, $B_0$ ($10^{9}$ Gauss) \dotfill & \multicolumn{2}{c}{2.9}\\
  Characteristic Age, $\tau_c$ (Gyr) \dotfill & \multicolumn{2}{c}{4.1}\\
  \hline
  \tablecomments{Numbers in parentheses represent 1-$\sigma$
    uncertainties in the last digits as determined by \texttt{TEMPO},
    scaled such that the reduced $\chi^2= 1$. Note that the timing
    parameters are given in two different timescales, TDB and TCB.
  (d) indicates a parameter that is derived in one model, but fitted
    directly in the other. (m) Parameter derived
    from the Bayesian analysis described in the text.
    The distance is derived from the DM using the \cite{CordesLazio2002}
    model of the Galactic electron density with a $\sim$25\% uncertainty.\nocite{CordesLazio2002}
  }
\end{tabular} }
\end{center}
\label{table:0453}
\end{table}

\section{Results}

The pulsar's ephemeris in Table~\ref{table:0453} includes a precise
position in the sky, which allows for optical follow-up. No optical
counterpart to the system is detectable in the online DSS2 optical
survey, either in the red or blue filters, nor in the 2MASS survey.
 
We can detect the proper motion in RA, and have useful limits
for the proper motion in Declination. We can derive a total proper motion
$\mu\, =\, (7.0\, \pm\, 2.5)\, \rm mas \, yr^{-1}$. The ephemeris also includes the
pulsar's spin period ($P$) and its derivative ($\dot{P}$). Taking into account the
effect of the proper motion (\citealt{Shklovskii70})
and Galactic acceleration (\citealt{DamourTaylor91})
on $\dot{P}$ at the DM-derived distance of 1.1 kpc, we obtain an
intrinsic $\dot{P}$ of 1.77(2)$\times$10$^{-19}$${\rm ~s~s^{-1}}$. From this
we derive a characteristic age of $4.1\, \times \, 10^9$ years
and a surface inferred magnetic field of $2.9 \, \times \, 10^{9}$ G. These
numbers are similar to what we observe for other recycled pulsars with
massive companions, and they indicate that this pulsar was mildly
recycled by accretion of matter from the progenitor of the current
companion star (\citealt{TaurisLangerKramer}).

The ephemeris also includes very precise orbital parameters. The
orbital period $P_b$ is 4.07 days, i.e., this is not a tight system where we
might be able to measure the remaining PK parameters precisely: the Einstein
delay ($\gamma$) will take a long time to measure and it
will correlate strongly with the kinematic $\dot{x}$
term that arises from the proper motion
(\citealt{Arzoumanian96} \& \citealt{Kopeokin96ProperMotion});
the orbital decay due to the emission of
gravitational waves ($\dot{P}_{\rm b}$) will be extremely small
and masked by much larger kinematic contributions
(\citealt{Shklovskii70} \& \citealt{DamourTaylor91}). The projected semi-major
axis of the pulsar's orbit $x$ is 14.5 light seconds; from
this we can derive the Keplerian mass function:
\begin{equation}
f(M_{p},M_{c})  =  \frac{(M_{c} {\rm sin} i)^{3}}{ (M_{p} + M_{c})^{2}}
                =  \frac{4\pi^{2}x^{3}}{T_{\odot}P_{b}^{2}} = 0.1960128(2)\, \Msun,
\label{eq:massf}
\end{equation}
where $T_{\odot} = G \Msun c^{-3} = 4.925490947 \mu \rm s$ is a
solar mass ($\Msun$) in time units ($c$ is the speed of light and
$G$ is Newton's gravitational constant), $i$ is the angle between the
plane of the
orbit and the plane of the sky, and the
pulsar and companion masses $M_{p}$ and $M_{c}$, are in solar
masses. If we assume for the pulsar a mass of 1.4 $\Msun$ and maximum
and median orbital inclinations ($i = 90^\circ, 60^\circ$) we obtain
minimum and median companion masses of 1.05 and 1.30 $\Msun$, i.e.,
the companion is relatively massive.

Given the orbital eccentricity ($e = 0.11$),
the companion is very likely to be a neutron star -- if it had
evolved into a massive white dwarf star there would be no 
sudden mass loss associated with a supernova explosion and the
system would have retained the circular orbit that is
characteristic of compact accreting systems. This is consistent with
the non-detection of an optical counterpart of the system in any of the
optical catalogs. However, the high eccentricity does not entirely
settle the matter: the recent discovery of a recycled
pulsar with a massive ($\sim 1 M_{\odot}$) companion
PSR~J1727$-$2946 \cite{Lorimer15Ecc}
and an orbital eccentricity of 0.0456
bridges the previously observed eccentricity gap between systems with
NS and massive WD companions. 
In the remainder of this paper we will assume, with caution,
that the companion is a NS.

The pulsar's orbital eccentricity allows a detection of the
advance of periastron, $\dot{\omega}$. If both components are compact,
as implied by the optical non-detection, this is given by:
\begin{equation}
\dot{\omega} = \dot{\omega}_{\rm GR} + \dot{\omega}_{\rm K}.
\label{eq:Omegadot}
\end{equation}
The second term $\dot{\omega_{\rm K}}$ is caused by the changing of viewing geometry 
due to the proper motion $\mu$ (\citealt{DamourTaylor91}):
\begin{equation}
\left| \dot{\omega}_{\rm K} \right | = \left| \frac{\mu}{\sin i} \cos(\Theta_{\mu} - \Omega)\right |\, \leq \, 3.65 \times 10^{-6} \, \rm deg\,\,  yr^{-1},
\label{eq:OmegaK}
\end{equation}
where $\Theta_{\mu}$ is the position angle (PA) of the proper motion and $\Omega$ is the PA
of the line of nodes (the intersection of the orbital plane with the plane of the sky).
This term is currently one order of magnitude smaller than the experimental uncertainty in the
measurement of $\dot{\omega}$. Thus $\dot{\omega}_{\rm GR} \simeq \dot{\omega}$.

The first term depends only on the Keplerian orbital
parameters, which are already known precisely, and the total mass of the binary $M$.
Thus $M$ can be derived from a measurement of $\dot{\omega}_{\rm GR} $ (\citealt{WeisbergTaylorGRPSR}):
\begin{equation}
M\,=\,\frac{1}{T_{\odot}} \left[ \frac{\dot{\omega}_{\rm GR}}{3} (1- e^2)
  \right]^{\frac{3}{2}} \left( \frac{P_{\mathrm{b}}}{2\pi}
\right)^{\frac{5}{2}}.
\label{eq:M}
\end{equation}
This yields $M \, = 2.734 \, \pm \, 0.003 \, \Msun$. As we can see in
Table~\ref{table:DNSlist}, this is within the mass range of currently
known DNS systems. 

We also measure the Shapiro delay with some precision: In the DDH
solution $h_3$ and $\varsigma$ are measured with 10 and 17-$\sigma$
significance. This in principle allows a measurement of the system
masses using the Shapiro delay alone.

\begin{figure}[h!]
\label{figure:mass_mass}
\includegraphics[width=7in]{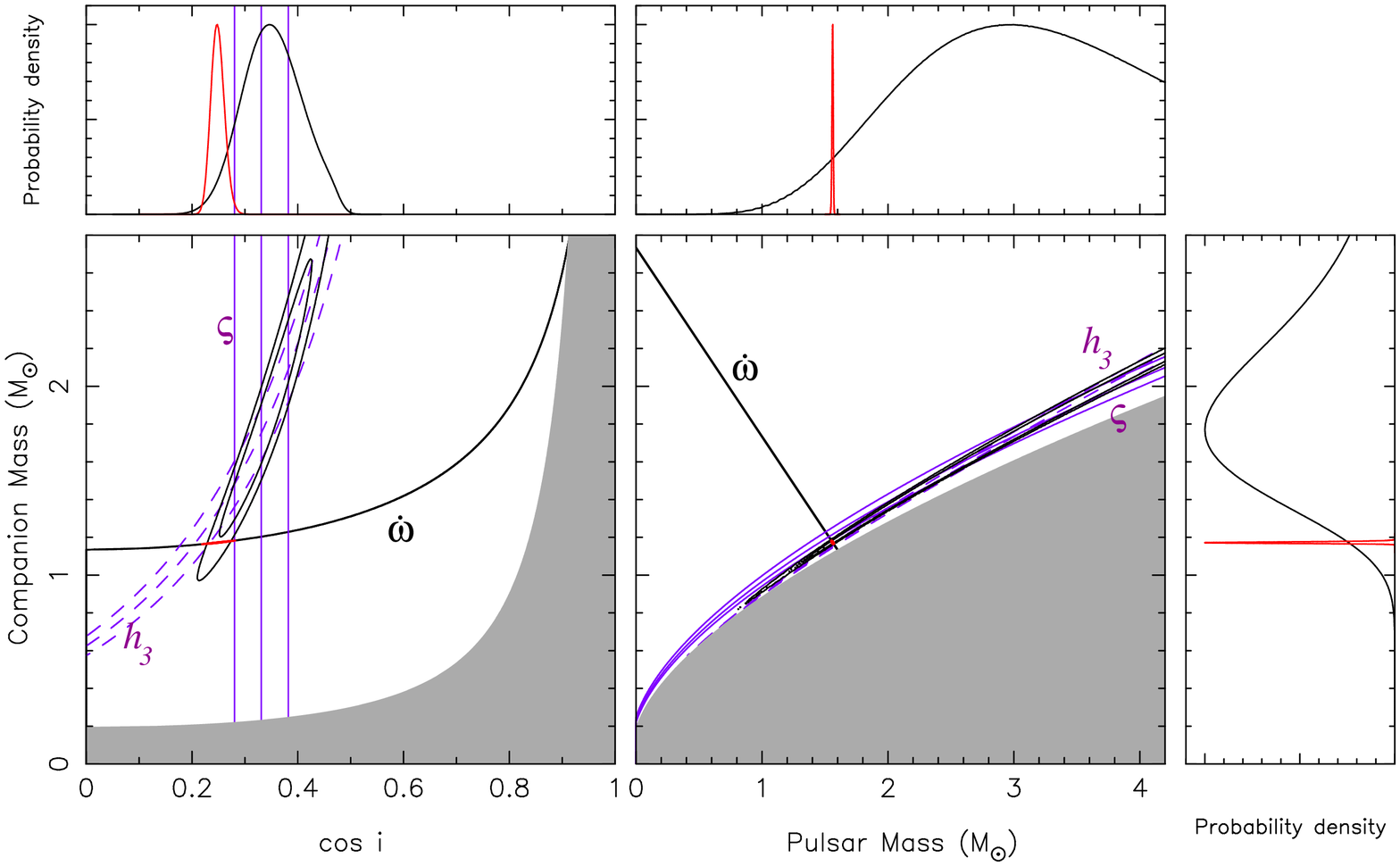}
\caption [] {\small Current constraints from timing of PSR
  J0453+1559. Each triplet of lines corresponds to the nominal and
  $\pm 1 \sigma$ uncertainties of the post-Keplerian parameters
  measured using the DDH model in {\sc tempo2} (see Table~\ref{table:0453}), 
  which are: the rate of advance of periastron $\dot{\omega}$, the
  orthometric ratio of the Shapiro delay $\varsigma$ and the orthometric
  amplitude of the Shapiro delay, $h_3$ \cite{FreireWex}.
  The contour levels contain 68.27 and 95.45\% of the 2-D probability
  density functions (pdfs) derived from the quality of the timing solution
  at each point of the $M_c$ - $\cos i$ plane using only the Shapiro
  delay (black) and Shapiro delay plus the assumption that
  the $\dot{\omega}$ is due only to the effects of GR (red).
  {\em Left}: $M_c$ - $\cos i$ plane. The gray region is excluded by
  the physical constraint $M_p > 0$. {\em Right}: $M_c$ - $M_p$
  plane. The gray region is excluded by the mathematical constraint
  $\sin i \leq 1$. {\em Top and right panels}: Pdfs for $\cos i$, $M_p$
  and (on the right) $M_c$, derived from
  marginalizing the 2-D pdf in the main panel for these
  quantities. When $\dot{\omega}$ is taken into account (red), the
  precision of the mass estimates improves by two orders of magnitude.
}
\end{figure}

In order to estimate the masses from the Shapiro delay, we sampled the
quality of the fit for a wide region in the $M_c - \cos i$ plane,
depicted in the left panel of Figure~\ref{figure:mass_mass}. For
each point in this plane, we calculate the Shapiro delay parameters
assuming GR to be the correct theory of gravity and then introduce
them in the timing solution, keeping them fixed and fitting for all
other timing parameters. The quality of the fit is quantified by the
$\chi^2$ of the resulting solution, the lower this is the better the
fit. From this $\chi^2$ map we derive a 2-D probability distribution 
function (pdf) using the Bayesian specification in \cite{Splaver+02}.
This is then converted to a similar 2-D pdf in the $M_c - M_p$ plane
(right panel) using eq.~\ref{eq:massf}. 
The black contours of both panels
include 68.27 and 95.45\% of the total probability of each pdf.
We then marginalize the 2-D pdfs to derive 1-D pdfs for $M_c$, $\cos i$
and $M_p$, the latter are presented in the
top and right panels in black. They allow for a wide (and rather
un-interesting) range of masses for the pulsar and the companion, i.e.,
just by itself the Shapiro delay does not provide useful mass constraints.

The precision of these mass estimates increases by two orders of 
magnitude if for each point in the $M_c - \cos i$ plane we fix
the Shapiro delay parameters {\em and} assume that $\dot{\omega}$
is due only to  the effects of GR (using eq.~\ref{eq:M}). As before,
we then fit for all other timing parameters, store the values of $\chi^2$
and calculate a second 2-D pdf. The latter is illustrated in
Figure~\ref{figure:mass_mass} by the red regions, which include
95.4\% of its total probability. 
Marginalizing this second pdf along the different axes we
obtain for the medians and 1-$\sigma$ percentiles the following values:
$M_{p} \, = \, 1.559\, \pm 0.005 \, \Msun$, $M_{c} \, = \, 1.172 \, \pm \, 0.004
\, \Msun$ and $i = 75.7^{+0.7}_{-0.8}$$^\circ$, respectively. This is
consistent with the masses provided by the DDGR model in {\tt TEMPO}.\\

\subsection{Search for the companion as a radio pulsar}
\label{sec:companion_search}

In order to search for radio pulsations from the companion, 
we used the early observations, which were taken in search mode. 
The precise measurement of the masses of the two NSs in the
system allows us to derive a complete ephemeris for the companion 
NS, except for the spin parameters and $\omega$, to which we add 180$^\circ$.
This ephemeris was used to 
resample the time series (dedispersed at the DM of the known pulsar) 
for the reference frame of the companion. This way, we
removed any possible losses in sensitivity due to the companion acceleration.
We then used the PRESTO pulsar search
code\footnote{http://www.cv.nrao.edu/$\sim$sransom/presto/} to search for periodic signals
in the resampled time series, and used it to fold all the candidates, which were
then inspected visually. No pulsations coming from the companion were detected.

To estimate the upper limit on the companion's pulsed mean flux density in our line of sight, 
$S_{\rm max}$, we used the radiometer equation (\citealt{handbook}), using the
parameters of the search observations and accounting for dispersive smearing at the
DM of the system and a minimum signal-to-noise ratio of 10. For L-band
observations the parameters are those reported in Section 2, whereas at 327 MHz
we have $T_{\rm sys} = 113$ K, $G=11$ K/Jy, $\Delta f = 60$ MHz.
At L-band, the longest observation had a length of $t_{\rm obs}=6300$ s, whereas
for only 327-MHz observation available to us $t_{\rm obs}=240$ s.

\begin{figure}[!h]
\includegraphics[width=\textwidth]{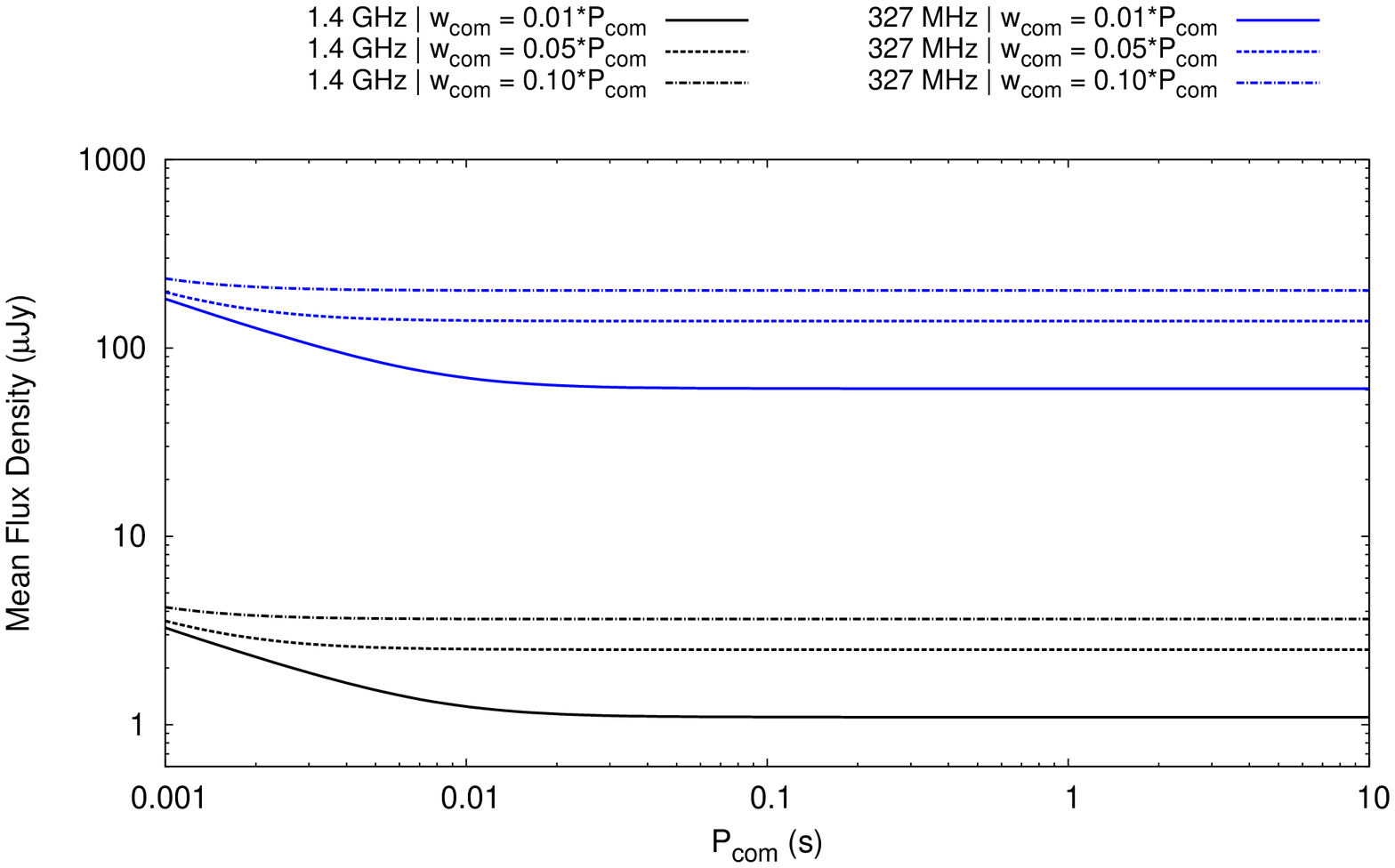}
\caption{Estimated minimum mean flux density of the companion if detectable with a S/N = 10, as a function of its spin period $P_{\textrm{com}}$ at 1.4 GHz (black lines) and 327 MHz (blue lines). For each frequency the cases of an intrinsic duty cycle $w_{\rm{com} }$ of 1\% (solid line), 5\% (dashed line) and 10\% (dot-dashed line) of $P_{\textrm{com}}$ are shown. At L-band, the parameters used were $t_{\rm obs} = 6300$ s, $T_{\rm sys} = 30$ K, $G=10$ K/Jy, $\Delta f = 600$ MHz; at 327 MHz they were $t_{\rm obs} = 240$ s, $T_{\rm sys} = 113$ K, $G=11$ K/Jy, $\Delta f = 60$ MHz; at both frequencies $\beta = 1$ and $n_p = 2$. Because the companion was not detected, its mean flux density must be towards out of our line of sight.}
\label{fig:sensitivitycurves}
\end{figure}

Figure \ref{fig:sensitivitycurves} shows $S_{\rm max}$ as a function of
the unknown companion spin period, $P_{\textrm{com}}$,
for intrinsic duty cycles of 1, 5 and 10\%. For the range of expected spin
periods of the companion ($P_{\textrm{com}} \gtrsim 0.1$ s),
$S_{\rm max, 1400} \lesssim 4\, \mu$Jy and $S_{\rm max, 327} \lesssim 202\, \mu$Jy.
Given the estimated distance of 1.0 kpc, these translate into pseudo-luminosities
of $L_{\rm max, 1400} \lesssim 4 \,  \mu$Jy kpc$^2$ and
$L_{\rm max, 327} \lesssim 202\, \mu$Jy kpc$^2$, respectively. No pulsar in the
ATNF catalog\footnote{http://www.atnf.csiro.au/people/pulsar/psrcat/}
has an estimated $L_{1400}$ as low as this.

\section{Discussion and conclusions}

The accretion episode in double neutron star systems is very short
lived and therefore the mass of the recycled pulsar is only slightly
larger than its mass at birth (\citealt{TaurisLangerPodsiadlowski15}).  
Until now, most well measured NS masses in DNS 
systems fell on a narrow range between 1.23 and 1.44 \Msun
(\citealt{Weisberg} and \citealt{J1756}, see Table~\ref {table:DNSlist}).
This has led to speculation that all NSs might be born within this
narrow band, and that the large masses observed in some MSPs like PSR
J1903+0327 (\citealt{FreireJ1903}), PSR~J1614$-$2230
(\citealt{Demorest2M}) and PSR~J0348+0432 (\citealt{Antoniadis}) are
due to accretion. However, from an analysis of the evolution of
PSR~J1614$-$2230, \citealt{TaurisKramer} had already suggested that at
least some NSs must be born more massive than 1.44 \Msun. The
mass of PSR~J0453+1559 - the largest ever measured in a DNS system
(see Table~\ref {table:DNSlist}) - and that of its companion - the
smallest precisely measured for {\em any} NS - shows that the range
of NS birth masses is indeed substantially wider than earlier studies
indicated (\citealt{Thorsett99} \& \citealt{Ozel12}). 

It is interesting to speculate on how the companion might have formed.
Its mass is lower than the 1.24 M$_{\odot}$
measured for the companion of J1756$-$2251 (\citealt{Ferdman14}) and PSRs J1802$-$2124 (\citealt{Ferdman10})
and J0737$-$3039B (\citealt{GRTestsdoublePSR}) that are thought to have
formed in electron capture supernovae (ECSN). It is possible that the companion
formed instead in an iron core collapse SN, where the core of the progenitor
of the companion was stripped of its envelope (Tauris et al., 2015, in preparation).

The relatively small eccentricity of the system compared 
to other DNS systems (see Table~\ref {table:DNSlist}) suggests a 
relatively small SN kick velocity at birth, which is consistent both with
formation via ECSN and an ultra-stripped iron core SN. In
this case the 3-D velocity of the system in the Galaxy
should be small in comparison with the general
pulsar population. This is consistent with the inferred
transverse velocity of $\sim$ 40 km s$^{-1}$ and the
relatively small Galactic height of the system, 0.29 kpc. 

The mass asymmetry is also important because it leads to a peculiar
behavior during NS-NS mergers, particularly if the mass ratio is less
than 0.8. During the merger the lighter (and larger)  
NS is tidally disrupted by the smaller, more massive NS. According to
recent simulations (\citealt{Rezzolla}, \citealt{Hotokezaka}
and \cite{Rosswog13}), such
mergers result in a much larger release of heavy r-process elements to space 
(\citealt{Just15}),
possibly explaining the heavy element abundances in our Galaxy.
However, asymmetric DNSs can only be an explanation for heavy element
abundances if they form with an orbital period that is small enough
for them to merge well within a Hubble time. Pulsar J0453+1559
has a very large merger time, 1.43 Tyr. However, the mass asymmetry
measured in this system opens up the possibility that
similar asymmetries might eventually be measured for compact DNSs
in the future.

This result is also important for searches of gravitational wave
emission from NS-NS mergers using ground-based GW detectors (\citealt{GWsDetectors1} \& \citealt{GWsDetectors2}) 
- it shows that we should not assume that the components of DNS systems 
have similar masses. This result justifies particularly searching for
DNS systems with lighter NSs, where more computational effort is required,
since in this case the inspiral episodes leading to the merger are 
significantly longer.

In asymmetric DNSs dipolar gravitational wave emission could
theoretically become important at the later stages of the merger;
however this possibility is already significantly constrained by the
measurement of the orbital decay of PSR~J1738+0333
(\citealt{Freire+12}) and J0348+0432
(\citealt{Antoniadis}), at least in the framework of
Scalar-Tensor theories of gravity. Therefore, GR-derived templates should be a
satisfactory approximation to the merger signal of asymmetric DNSs.

\section*{Acknowledgments}
J.S.D. was supported by the Chief of Naval Research. P.C.C.F. and A.R.
gratefully acknowledges financial support by the European Research 
Council for the ERC Starting Grant BEACON under contract no.~279702. 
J.G.M. was supported for this research through a stipend from the 
International Max Planck Research School (IMPRS) for Astronomy and Astrophysics at the Universities of Bonn and Cologne. A. R. is a member of the International Max Planck research school for Astronomy and Astrophysics at the Universities of Bonn and Cologne and acknowledges partial support through the Bonn-Cologne Graduate School of Physics and Astronomy.

This work would have not been possible without the high sensitivity of the Arecibo 305-m radio telescope, the professionalism and dedication of its excellent staff and the capabilities of the PUPPI back-end, a clone of the GUPPI developed by NRAO and kindly made available to the whole community by its developers, Scott Ransom, Paul Demorest and the development team at Charlottesville, VA. The Arecibo Observatory is operated by SRI International under a cooperative agreement with the National Science Foundation (AST-1100968), and in alliance with Ana G. M\'endez- Universidad Metropolitana, and the Universities Space Research Association. 

J.G.M. would like to deeply thank the Arecibo Remote 
Command Center (ARCC) program in the Center for Advanced Radio Astronomy (CARA) 
center at the University of Texas at Brownsville. This program provided the 
the setting and support that allowed this work to be possible. Thank you to the ARCC Executive Committee for your dedication and hard work.

\bibliographystyle{apj}
\bibliography{ref}

\begin{thebibliography}{57}
\expandafter\ifx\csname natexlab\endcsname\relax\def\natexlab#1{#1}\fi

\bibitem[{{Abadie} {et~al.}(2012){Abadie}, {Abbott}, {Abbott}, {Abbott},
  {Abernathy}, {Accadia}, {Acernese}, {Adams}, {Adhikari}, {Affeldt}, \&
  et~al.}]{GWsDetectors2}
{Abadie}, J., {Abbott}, B.~P., {Abbott}, R., {Abbott}, T.~D., {Abernathy}, M.,
  {Accadia}, T., {Acernese}, F., {Adams}, C., {Adhikari}, R., {Affeldt}, C., \&
  et~al. 2012, \prd, 85, 082002

\bibitem[{{Abbott} {et~al.}(2009){Abbott}, {Abbott}, {Adhikari}, {Ajith},
  {Allen}, {Allen}, {Amin}, {Anderson}, {Anderson}, {Arain}, \&
  et~al.}]{GWsDetectors1}
{Abbott}, B.~P., {Abbott}, R., {Adhikari}, R., {Ajith}, P., {Allen}, B.,
  {Allen}, G., {Amin}, R.~S., {Anderson}, S.~B., {Anderson}, W.~G., {Arain},
  M.~A., \& et~al. 2009, \prd, 79, 122001

\bibitem[{{Anderson} {et~al.}(1989){Anderson}, {Gorham}, {Kulkarni}, {Prince},
  \& {Wolszczan}}]{B2127}
{Anderson}, S., {Gorham}, P., {Kulkarni}, S., {Prince}, T., \& {Wolszczan}, A.
  1989, \iaucirc, 4772, 1

\bibitem[{{Antoniadis} {et~al.}(2013){Antoniadis}, {Freire}, {Wex}, {Tauris},
  {Lynch}, {van Kerkwijk}, {Kramer}, {Bassa}, {Dhillon}, {Driebe}, {Hessels},
  {Kaspi}, {Kondratiev}, {Langer}, {Marsh}, {McLaughlin}, {Pennucci}, {Ransom},
  {Stairs}, {van Leeuwen}, {Verbiest}, \& {Whelan}}]{Antoniadis}
{Antoniadis}, J., {Freire}, P.~C.~C., {Wex}, N., {Tauris}, T.~M., {Lynch},
  R.~S., {van Kerkwijk}, M.~H., {Kramer}, M., {Bassa}, C., {Dhillon}, V.~S.,
  {Driebe}, T., {Hessels}, J.~W.~T., {Kaspi}, V.~M., {Kondratiev}, V.~I.,
  {Langer}, N., {Marsh}, T.~R., {McLaughlin}, M.~A., {Pennucci}, T.~T.,
  {Ransom}, S.~M., {Stairs}, I.~H., {van Leeuwen}, J., {Verbiest}, J.~P.~W., \&
  {Whelan}, D.~G. 2013, Science, 340, 448

\bibitem[{{Arzoumanian} {et~al.}(1996){Arzoumanian}, {Joshi}, {Rasio}, \&
  {Thorsett}}]{Arzoumanian96}
{Arzoumanian}, Z., {Joshi}, K., {Rasio}, F.~A., \& {Thorsett}, S.~E. 1996, in
  Astronomical Society of the Pacific Conference Series, Vol. 105, IAU Colloq.
  160: Pulsars: Problems and Progress, ed. S.~{Johnston}, M.~A. {Walker}, \&
  M.~{Bailes}, 525--530

\bibitem[{{Burgay} {et~al.}(2003){Burgay}, {D'Amico}, {Possenti}, {Manchester},
  {Lyne}, {Joshi}, {McLaughlin}, {Kramer}, {Sarkissian}, {Camilo}, {Kalogera},
  {Kim}, \& {Lorimer}}]{doublePSR}
{Burgay}, M., {D'Amico}, N., {Possenti}, A., {Manchester}, R.~N., {Lyne},
  A.~G., {Joshi}, B.~C., {McLaughlin}, M.~A., {Kramer}, M., {Sarkissian},
  J.~M., {Camilo}, F., {Kalogera}, V., {Kim}, C., \& {Lorimer}, D.~R. 2003,
  \nat, 426, 531

\bibitem[{{Champion} {et~al.}(2004){Champion}, {Lorimer}, {McLaughlin},
  {Cordes}, {Arzoumanian}, {Weisberg}, \& {Taylor}}]{J1829}
{Champion}, D.~J., {Lorimer}, D.~R., {McLaughlin}, M.~A., {Cordes}, J.~M.,
  {Arzoumanian}, Z., {Weisberg}, J.~M., \& {Taylor}, J.~H. 2004, \mnras, 350,
  L61

\bibitem[{{Champion} {et~al.}(2005){Champion}, {Lorimer}, {McLaughlin},
  {Xilouris}, {Arzoumanian}, {Freire}, {Lommen}, {Cordes}, \&
  {Camilo}}]{J1829_2}
{Champion}, D.~J., {Lorimer}, D.~R., {McLaughlin}, M.~A., {Xilouris}, K.~M.,
  {Arzoumanian}, Z., {Freire}, P.~C.~C., {Lommen}, A.~N., {Cordes}, J.~M., \&
  {Camilo}, F. 2005, \mnras, 363, 929

\bibitem[{{Cordes} \& {Lazio}(2002)}]{CordesLazio2002}
{Cordes}, J.~M., \& {Lazio}, T.~J.~W. 2002, ArXiv Astrophysics e-prints

\bibitem[{{Corongiu} {et~al.}(2007){Corongiu}, {Kramer}, {Stappers}, {Lyne},
  {Jessner}, {Possenti}, {D'Amico}, \& {L{\"o}hmer}}]{CorongiuJ1811-1736}
{Corongiu}, A., {Kramer}, M., {Stappers}, B.~W., {Lyne}, A.~G., {Jessner}, A.,
  {Possenti}, A., {D'Amico}, N., \& {L{\"o}hmer}, O. 2007, \aap, 462, 703

\bibitem[{{Damour} \& {Deruelle}(1985)}]{Damour85}
{Damour}, T., \& {Deruelle}, N. 1985, Ann.~Inst.~Henri Poincar{\'e}
  Phys.~Th{\'e}or., Vol.~43, No.~1, p.~107 - 132, 43, 107

\bibitem[{{Damour} \& {Deruelle}(1986)}]{Damour86}
---. 1986, Ann.~Inst.~Henri Poincar{\'e} Phys.~Th{\'e}or., Vol.~44, No.~3,
  p.~263 - 292, 44, 263

\bibitem[{{Damour} \& {Taylor}(1991)}]{DamourTaylor91}
{Damour}, T., \& {Taylor}, J.~H. 1991, \apj, 366, 501

\bibitem[{{Demorest} {et~al.}(2010){Demorest}, {Pennucci}, {Ransom}, {Roberts},
  \& {Hessels}}]{Demorest2M}
{Demorest}, P.~B., {Pennucci}, T., {Ransom}, S.~M., {Roberts}, M.~S.~E., \&
  {Hessels}, J.~W.~T. 2010, \nat, 467, 1081

\bibitem[{{Deneva} {et~al.}(2013){Deneva}, {Stovall}, {McLaughlin}, {Bates},
  {Freire}, {Martinez}, {Jenet}, \& {Bagchi}}]{DenevaAODrift}
{Deneva}, J.~S., {Stovall}, K., {McLaughlin}, M.~A., {Bates}, S.~D., {Freire},
  P.~C.~C., {Martinez}, J.~G., {Jenet}, F., \& {Bagchi}, M. 2013, \apj, 775, 51

\bibitem[{{Edwards} {et~al.}(2006){Edwards}, {Hobbs}, \&
  {Manchester}}]{tempo22}
{Edwards}, R.~T., {Hobbs}, G.~B., \& {Manchester}, R.~N. 2006, \mnras, 372,
  1549

\bibitem[{{Faulkner} {et~al.}(2005){Faulkner}, {Kramer}, {Lyne}, {Manchester},
  {McLaughlin}, {Stairs}, {Hobbs}, {Possenti}, {Lorimer}, {D'Amico}, {Camilo},
  \& {Burgay}}]{J1756}
{Faulkner}, A.~J., {Kramer}, M., {Lyne}, A.~G., {Manchester}, R.~N.,
  {McLaughlin}, M.~A., {Stairs}, I.~H., {Hobbs}, G., {Possenti}, A., {Lorimer},
  D.~R., {D'Amico}, N., {Camilo}, F., \& {Burgay}, M. 2005, \apjl, 618, L119

\bibitem[{{Ferdman} {et~al.}(2014{\natexlab{a}}){Ferdman}, {Stairs}, {Kramer},
  {Janssen}, {Bassa}, {Stappers}, {Demorest}, {Cognard}, {Desvignes},
  {Theureau}, {Burgay}, {Lyne}, {Manchester}, \& {Possenti}}]{J1756_updated}
{Ferdman}, R.~D., {Stairs}, I.~H., {Kramer}, M., {Janssen}, G.~H., {Bassa},
  C.~G., {Stappers}, B.~W., {Demorest}, P.~B., {Cognard}, I., {Desvignes}, G.,
  {Theureau}, G., {Burgay}, M., {Lyne}, A.~G., {Manchester}, R.~N., \&
  {Possenti}, A. 2014{\natexlab{a}}, \mnras, 443, 2183

\bibitem[{{Ferdman} {et~al.}(2014{\natexlab{b}}){Ferdman}, {Stairs}, {Kramer},
  {Janssen}, {Bassa}, {Stappers}, {Demorest}, {Cognard}, {Desvignes},
  {Theureau}, {Burgay}, {Lyne}, {Manchester}, \& {Possenti}}]{Ferdman14}
---. 2014{\natexlab{b}}, \mnras, 443, 2183

\bibitem[{{Ferdman} {et~al.}(2010){Ferdman}, {Stairs}, {Kramer}, {McLaughlin},
  {Lorimer}, {Nice}, {Manchester}, {Hobbs}, {Lyne}, {Camilo}, {Possenti},
  {Demorest}, {Cognard}, {Desvignes}, {Theureau}, {Faulkner}, \&
  {Backer}}]{Ferdman10}
{Ferdman}, R.~D., {Stairs}, I.~H., {Kramer}, M., {McLaughlin}, M.~A.,
  {Lorimer}, D.~R., {Nice}, D.~J., {Manchester}, R.~N., {Hobbs}, G., {Lyne},
  A.~G., {Camilo}, F., {Possenti}, A., {Demorest}, P.~B., {Cognard}, I.,
  {Desvignes}, G., {Theureau}, G., {Faulkner}, A., \& {Backer}, D.~C. 2010,
  \apj, 711, 764

\bibitem[{{Fonseca} {et~al.}(2014){Fonseca}, {Stairs}, \&
  {Thorsett}}]{B1534_updated}
{Fonseca}, E., {Stairs}, I.~H., \& {Thorsett}, S.~E. 2014, \apj, 787, 82

\bibitem[{{Freire} {et~al.}(2011){Freire}, {Bassa}, {Wex}, {Stairs},
  {Champion}, {Ransom}, {Lazarus}, {Kaspi}, {Hessels}, {Kramer}, {Cordes},
  {Verbiest}, {Podsiadlowski}, {Nice}, {Deneva}, {Lorimer}, {Stappers},
  {McLaughlin}, \& {Camilo}}]{FreireJ1903}
{Freire}, P.~C.~C., {Bassa}, C.~G., {Wex}, N., {Stairs}, I.~H., {Champion},
  D.~J., {Ransom}, S.~M., {Lazarus}, P., {Kaspi}, V.~M., {Hessels}, J.~W.~T.,
  {Kramer}, M., {Cordes}, J.~M., {Verbiest}, J.~P.~W., {Podsiadlowski}, P.,
  {Nice}, D.~J., {Deneva}, J.~S., {Lorimer}, D.~R., {Stappers}, B.~W.,
  {McLaughlin}, M.~A., \& {Camilo}, F. 2011, \mnras, 412, 2763

\bibitem[{{Freire} \& {Wex}(2010)}]{FreireWex}
{Freire}, P.~C.~C., \& {Wex}, N. 2010, \mnras, 409, 199

\bibitem[{{Freire} {et~al.}(2012){Freire}, {Wex}, {Esposito-Far{\`e}se},
  {Verbiest}, {Bailes}, {Jacoby}, {Kramer}, {Stairs}, {Antoniadis}, \&
  {Janssen}}]{Freire+12}
{Freire}, P.~C.~C., {Wex}, N., {Esposito-Far{\`e}se}, G., {Verbiest}, J.~P.~W.,
  {Bailes}, M., {Jacoby}, B.~A., {Kramer}, M., {Stairs}, I.~H., {Antoniadis},
  J., \& {Janssen}, G.~H. 2012, \mnras, 423, 3328

\bibitem[{{Hobbs} {et~al.}(2006){Hobbs}, {Edwards}, \& {Manchester}}]{tempo2}
{Hobbs}, G.~B., {Edwards}, R.~T., \& {Manchester}, R.~N. 2006, \mnras, 369, 655

\bibitem[{{Hotan} {et~al.}(2004){Hotan}, {van Straten}, \&
  {Manchester}}]{PSRCHIVE}
{Hotan}, A.~W., {van Straten}, W., \& {Manchester}, R.~N. 2004, \pasa, 21, 302

\bibitem[{{Hotokezaka} {et~al.}(2013){Hotokezaka}, {Kiuchi}, {Kyutoku},
  {Okawa}, {Sekiguchi}, {Shibata}, \& {Taniguchi}}]{Hotokezaka}
{Hotokezaka}, K., {Kiuchi}, K., {Kyutoku}, K., {Okawa}, H., {Sekiguchi}, Y.-i.,
  {Shibata}, M., \& {Taniguchi}, K. 2013, \prd, 87, 024001

\bibitem[{{Hulse} \& {Taylor}(1975)}]{ht75}
{Hulse}, R.~A., \& {Taylor}, J.~H. 1975, \apjl, 195, L51

\bibitem[{{Jacoby} {et~al.}(2006){Jacoby}, {Cameron}, {Jenet}, {Anderson},
  {Murty}, \& {Kulkarni}}]{B2127_updated}
{Jacoby}, B.~A., {Cameron}, P.~B., {Jenet}, F.~A., {Anderson}, S.~B., {Murty},
  R.~N., \& {Kulkarni}, S.~R. 2006, \apjl, 644, L113

\bibitem[{{Janssen} {et~al.}(2008){Janssen}, {Stappers}, {Kramer}, {Nice},
  {Jessner}, {Cognard}, \& {Purver}}]{JanssenJ1518+4904}
{Janssen}, G.~H., {Stappers}, B.~W., {Kramer}, M., {Nice}, D.~J., {Jessner},
  A., {Cognard}, I., \& {Purver}, M.~B. 2008, \aap, 490, 753

\bibitem[{{Just} {et~al.}(2015){Just}, {Bauswein}, {Ardevol Pulpillo},
  {Goriely}, \& {Janka}}]{Just15}
{Just}, O., {Bauswein}, A., {Ardevol Pulpillo}, R., {Goriely}, S., \& {Janka},
  H.-T. 2015, ArXiv e-prints

\bibitem[{{Keith} {et~al.}(2009){Keith}, {Kramer}, {Lyne}, {Eatough}, {Stairs},
  {Possenti}, {Camilo}, \& {Manchester}}]{J1753}
{Keith}, M.~J., {Kramer}, M., {Lyne}, A.~G., {Eatough}, R.~P., {Stairs}, I.~H.,
  {Possenti}, A., {Camilo}, F., \& {Manchester}, R.~N. 2009, \mnras, 393, 623

\bibitem[{{Kopeikin}(1996)}]{Kopeokin96ProperMotion}
{Kopeikin}, S.~M. 1996, \apjl, 467, L93

\bibitem[{{Kramer} {et~al.}(2006){Kramer}, {Stairs}, {Manchester},
  {McLaughlin}, {Lyne}, {Ferdman}, {Burgay}, {Lorimer}, {Possenti}, {D'Amico},
  {Sarkissian}, {Hobbs}, {Reynolds}, {Freire}, \& {Camilo}}]{GRTestsdoublePSR}
{Kramer}, M., {Stairs}, I.~H., {Manchester}, R.~N., {McLaughlin}, M.~A.,
  {Lyne}, A.~G., {Ferdman}, R.~D., {Burgay}, M., {Lorimer}, D.~R., {Possenti},
  A., {D'Amico}, N., {Sarkissian}, J.~M., {Hobbs}, G.~B., {Reynolds}, J.~E.,
  {Freire}, P.~C.~C., \& {Camilo}, F. 2006, Science, 314, 97

\bibitem[{{Lorimer}(2008)}]{DuncReview}
{Lorimer}, D.~R. 2008, Living Reviews in Relativity, 11, 8

\bibitem[{{Lorimer} {et~al.}(2015){Lorimer}, {Esposito}, {Manchester},
  {Possenti}, {Lyne}, {McLaughlin}, {Kramer}, {Hobbs}, {Stairs}, {Burgay},
  {Eatough}, {Keith}, {Faulkner}, {D'Amico}, {Camilo}, {Corongiu}, \&
  {Crawford}}]{Lorimer15Ecc}
{Lorimer}, D.~R., {Esposito}, P., {Manchester}, R.~N., {Possenti}, A., {Lyne},
  A.~G., {McLaughlin}, M.~A., {Kramer}, M., {Hobbs}, G., {Stairs}, I.~H.,
  {Burgay}, M., {Eatough}, R.~P., {Keith}, M.~J., {Faulkner}, A.~J., {D'Amico},
  N., {Camilo}, F., {Corongiu}, A., \& {Crawford}, F. 2015, \mnras, 450, 2185

\bibitem[{{Lorimer} \& {Kramer}(2004)}]{handbook}
{Lorimer}, D.~R., \& {Kramer}, M. 2004, {Handbook of Pulsar Astronomy}

\bibitem[{{Lorimer} {et~al.}(2006){Lorimer}, {Stairs}, {Freire}, {Cordes},
  {Camilo}, {Faulkner}, {Lyne}, {Nice}, {Ransom}, {Arzoumanian}, {Manchester},
  {Champion}, {van Leeuwen}, {Mclaughlin}, {Ramachandran}, {Hessels},
  {Vlemmings}, {Deshpande}, {Bhat}, {Chatterjee}, {Han}, {Gaensler}, {Kasian},
  {Deneva}, {Reid}, {Lazio}, {Kaspi}, {Crawford}, {Lommen}, {Backer}, {Kramer},
  {Stappers}, {Hobbs}, {Possenti}, {D'Amico}, \& {Burgay}}]{J1906}
{Lorimer}, D.~R., {Stairs}, I.~H., {Freire}, P.~C., {Cordes}, J.~M., {Camilo},
  F., {Faulkner}, A.~J., {Lyne}, A.~G., {Nice}, D.~J., {Ransom}, S.~M.,
  {Arzoumanian}, Z., {Manchester}, R.~N., {Champion}, D.~J., {van Leeuwen}, J.,
  {Mclaughlin}, M.~A., {Ramachandran}, R., {Hessels}, J.~W., {Vlemmings}, W.,
  {Deshpande}, A.~A., {Bhat}, N.~D., {Chatterjee}, S., {Han}, J.~L.,
  {Gaensler}, B.~M., {Kasian}, L., {Deneva}, J.~S., {Reid}, B., {Lazio}, T.~J.,
  {Kaspi}, V.~M., {Crawford}, F., {Lommen}, A.~N., {Backer}, D.~C., {Kramer},
  M., {Stappers}, B.~W., {Hobbs}, G.~B., {Possenti}, A., {D'Amico}, N., \&
  {Burgay}, M. 2006, \apj, 640, 428

\bibitem[{{Lynch} {et~al.}(2012){Lynch}, {Freire}, {Ransom}, \&
  {Jacoby}}]{J1807}
{Lynch}, R.~S., {Freire}, P.~C.~C., {Ransom}, S.~M., \& {Jacoby}, B.~A. 2012,
  \apj, 745, 109

\bibitem[{{{\"O}zel} {et~al.}(2012){{\"O}zel}, {Psaltis}, {Narayan}, \& {Santos
  Villarreal}}]{Ozel12}
{{\"O}zel}, F., {Psaltis}, D., {Narayan}, R., \& {Santos Villarreal}, A. 2012,
  \apj, 757, 55

\bibitem[{{Rezzolla} {et~al.}(2010){Rezzolla}, {Baiotti}, {Giacomazzo}, {Link},
  \& {Font}}]{Rezzolla}
{Rezzolla}, L., {Baiotti}, L., {Giacomazzo}, B., {Link}, D., \& {Font}, J.~A.
  2010, Classical and Quantum Gravity, 27, 114105

\bibitem[{{Rosswog}(2013)}]{Rosswog13}
{Rosswog}, S. 2013, Royal Society of London Philosophical Transactions Series
  A, 371, 20272

\bibitem[{{Shklovskii}(1970)}]{Shklovskii70}
{Shklovskii}, I.~S. 1970, \sovast, 13, 562

\bibitem[{{Splaver} {et~al.}(2002){Splaver}, {Nice}, {Arzoumanian}, {Camilo},
  {Lyne}, \& {Stairs}}]{Splaver+02}
{Splaver}, E.~M., {Nice}, D.~J., {Arzoumanian}, Z., {Camilo}, F., {Lyne},
  A.~G., \& {Stairs}, I.~H. 2002, \apj, 581, 509

\bibitem[{{Swiggum} {et~al.}(2015){Swiggum}, {Rosen}, {McLaughlin}, {Lorimer},
  {Heatherly}, {Lynch}, {Scoles}, {Hockett}, {Filik}, {Marlowe}, {Barlow},
  {Weaver}, {Hilzendeger}, {Ernst}, {Crowley}, {Stone}, {Miller}, {Nunez},
  {Trevino}, {Doehler}, {Cramer}, {Yencsik}, {Thorley}, {Andrews}, {Laws},
  {Wenger}, {Teter}, {Snyder}, {Dittmann}, {Gray}, {Carter}, {McGough},
  {Dydiw}, {Pruett}, {Fink}, \& {Vanderhout}}]{J1930}
{Swiggum}, J.~K., {Rosen}, R., {McLaughlin}, M.~A., {Lorimer}, D.~R.,
  {Heatherly}, S., {Lynch}, R., {Scoles}, S., {Hockett}, T., {Filik}, E.,
  {Marlowe}, J.~A., {Barlow}, B.~N., {Weaver}, M., {Hilzendeger}, M., {Ernst},
  S., {Crowley}, R., {Stone}, E., {Miller}, B., {Nunez}, R., {Trevino}, G.,
  {Doehler}, M., {Cramer}, A., {Yencsik}, D., {Thorley}, J., {Andrews}, R.,
  {Laws}, A., {Wenger}, K., {Teter}, L., {Snyder}, T., {Dittmann}, A., {Gray},
  S., {Carter}, M., {McGough}, C., {Dydiw}, S., {Pruett}, C., {Fink}, J., \&
  {Vanderhout}, A. 2015, ArXiv e-prints

\bibitem[{{Tauris} {et~al.}(2011){Tauris}, {Langer}, \&
  {Kramer}}]{TaurisKramer}
{Tauris}, T.~M., {Langer}, N., \& {Kramer}, M. 2011, \mnras, 416, 2130

\bibitem[{{Tauris} {et~al.}(2012){Tauris}, {Langer}, \&
  {Kramer}}]{TaurisLangerKramer}
---. 2012, \mnras, 425, 1601

\bibitem[{{Tauris} {et~al.}(2015){Tauris}, {Langer}, \&
  {Podsiadlowski}}]{TaurisLangerPodsiadlowski15}
{Tauris}, T.~M., {Langer}, N., \& {Podsiadlowski}, P. 2015, ArXiv e-prints

\bibitem[{{Taylor}(1992)}]{TaylorRG}
{Taylor}, J.~H. 1992, Royal Society of London Philosophical Transactions Series
  A, 341, 117

\bibitem[{{Taylor} \& {Weisberg}(1982)}]{htNobel}
{Taylor}, J.~H., \& {Weisberg}, J.~M. 1982, \apj, 253, 908

\bibitem[{{Thorsett} \& {Chakrabarty}(1999)}]{Thorsett99}
{Thorsett}, S.~E., \& {Chakrabarty}, D. 1999, \apj, 512, 288

\bibitem[{{van Leeuwen} {et~al.}(2015){van Leeuwen}, {Kasian}, {Stairs},
  {Lorimer}, {Camilo}, {Chatterjee}, {Cognard}, {Desvignes}, {Freire},
  {Janssen}, {Kramer}, {Lyne}, {Nice}, {Ransom}, {Stappers}, \&
  {Weisberg}}]{J1906_updated}
{van Leeuwen}, J., {Kasian}, L., {Stairs}, I.~H., {Lorimer}, D.~R., {Camilo},
  F., {Chatterjee}, S., {Cognard}, I., {Desvignes}, G., {Freire}, P.~C.~C.,
  {Janssen}, G.~H., {Kramer}, M., {Lyne}, A.~G., {Nice}, D.~J., {Ransom},
  S.~M., {Stappers}, B.~W., \& {Weisberg}, J.~M. 2015, \apj, 798, 118

\bibitem[{{van Straten} \& {Bailes}(2011)}]{dspsr}
{van Straten}, W., \& {Bailes}, M. 2011, \pasa, 28, 1

\bibitem[{{van Straten} {et~al.}(2012){van Straten}, {Demorest}, \&
  {Oslowski}}]{PSRCHIVE2012}
{van Straten}, W., {Demorest}, P., \& {Oslowski}, S. 2012, Astronomical
  Research and Technology, 9, 237

\bibitem[{{Weisberg} {et~al.}(2010){Weisberg}, {Nice}, \& {Taylor}}]{Weisberg}
{Weisberg}, J.~M., {Nice}, D.~J., \& {Taylor}, J.~H. 2010, \apj, 722, 1030

\bibitem[{{Weisberg} {et~al.}(1981){Weisberg}, {Taylor}, \&
  {Fowler}}]{WeisbergTaylorGRPSR}
{Weisberg}, J.~M., {Taylor}, J.~H., \& {Fowler}, L.~A. 1981, Scientific
  American, 245, 74

\bibitem[{{Wolszczan}(1991)}]{B1534}
{Wolszczan}, A. 1991, \nat, 350, 688

\end{thebibliography}
\acknowledgments

\end{document}